\documentclass[a4paper, aps, prb, superscriptaddress, twocolumn, longbibliography]{revtex4-1}
\setcitestyle{numbers,square}

\usepackage{stmaryrd}
\usepackage{amsfonts}
\usepackage{amssymb}
\usepackage{bbold}
\usepackage{mathrsfs,amsmath}    
\usepackage{graphicx}   
\usepackage{verbatim}   
\usepackage[dvipsnames]{xcolor}     
\usepackage{subfigure} 
\usepackage{hyperref}  
\usepackage{braket}

\usepackage{multirow}

\usepackage{empheq}

\usepackage{hyperref}
\hypersetup{colorlinks,citecolor={blue},linkcolor={red},urlcolor={blue}}

\makeatletter

\newsavebox{\@brx}
\newcommand{\llangle}[1][]{\savebox{\@brx}{\(\m@th{#1\langle}\)}%
  \mathopen{\copy\@brx\kern-0.5\wd\@brx\usebox{\@brx}}}
\newcommand{\rrangle}[1][]{\savebox{\@brx}{\(\m@th{#1\rangle}\)}%
  \mathclose{\copy\@brx\kern-0.5\wd\@brx\usebox{\@brx}}}
\makeatother

\usepackage{enumitem}

\usepackage{bm}
\renewcommand{\vec}{\ensuremath{\bm}}

\usepackage{bm}
\renewcommand{\vec}{\ensuremath{\bm}}

\usepackage{amsmath,amsfonts,amssymb}
\usepackage{bm}

\usepackage{rotating}
\usepackage{tabularx}
\usepackage{dcolumn}
\usepackage{nicefrac}

\newcommand{\dy}[1]{\bar{\bar{#1}}}

\begin{document}

\title{Three dimensional vectorial imaging of surface phonons}
\author{Xiaoyan Li}
\affiliation{Laboratoire de Physique des Solides, Université Paris-Sud, CNRS-UMR 8502, Orsay 91405, France}
\author{Georg Haberfehlner}
\affiliation{Institute of Electron Microscopy and Nanoanalysis, Graz University of Technology, Steyrergasse 17, 8010 Graz, Austria}
\author{Ulrich Hohenester}
\affiliation{Institute of Physics, University of Graz, Universitätsplatz 5, 8010 Graz, Austria}
\author{Odile Stéphan}
\affiliation{Laboratoire de Physique des Solides, Université Paris-Sud, CNRS-UMR 8502, Orsay 91405, France}
\author{Gerald Kothleitner}
\affiliation{Institute of Electron Microscopy and Nanoanalysis, Graz University of Technology, Steyrergasse 17, 8010 Graz, Austria}
\affiliation{Graz Centre for Electron Microscopy, Steyrergasse 17, 8010 Graz, Austria}
\author{Mathieu Kociak}
\affiliation{Laboratoire de Physique des Solides, Université Paris-Sud, CNRS-UMR 8502, Orsay 91405, France}

\date{\today}

\begin{abstract}
While phonons and their related properties have been studied comprehensively in bulk materials, a thorough understanding of surface phonons for nanoscale objects remains elusive. Infra-red imaging methods with photons \cite{Knoll1999,Hillenbrand2002,Zhang2013,DeWilde2006} or electrons \cite{Krivanek2014,Lagos2017} exist, which can map these vibrational excitations down to the individual atom scale \cite{Zhang2013,venkatraman2019,Hage2020}. However, no technique so far was able to directly reveal the complete three-dimensional vectorial picture of the phononic electromagnetic density of states \cite{Lourenco-Martins2017} from a nanostructured object. Using a highly monochromated electron beam in a scanning transmission electron microscope, we could visualize  varying phonon signatures from nanoscale MgO cubes both as a function of the beam position, energy-loss and tilt angle. The vibrational responses were described in terms of well-defined eigenmodes and were used to tomographically reconstruct the phononic surface fields of the object, disclosing effects invisible by conventional mapping. Surface phonon applications \cite{Caldwell2015,Neubrech2008,NeunerIII2010,Autore2018} rely critically on the way the electromagnetic field is spatially and vectorially shaped, strongly influencing the optical, thermal and electronic behavior of the materials. Such 3D information therefore promises novel insights in nanoscale physical phenomena and is invaluable to the design and optimization of nanostructures for fascinating new uses \cite{NeunerIII2010,Tizei2020,Rez2016}.
\end{abstract}

\maketitle



Phonons are quanta of elementary vibrational motions in solids and their energy and momenta distributions directly determine many fundamental thermodynamic properties of solids, such as heat capacitance or heat conductivity. Closely linked to the local arrangement of atoms, interesting new properties arise when symmetries are broken and surfaces are created, making surface phonons becoming strongly dependent on the object geometry \cite{Kliewer,Lourenco-Martins2017}.

Surface phonons have recently attracted much attention because of their counter-intuitive physical properties and their promising applications in photonics and nanophotonics from  the mid-IR (3-8 $\mathrm{\mu m}$, 155-413 meV) up to the far-IR (15-1000 $\mathrm{\mu m}$, 	1.2-83 meV) \cite{Caldwell2015}. For example, materials such as SiC possess strong surface phonon resonances and may exhibit a highly coherent emission upon heating. This behaviour differs completely from the conventional incoherent black-body radiation \cite{Greffet2002}. At the nanoscale, two surfaces sensing each other’s near-field, may exhibit strongly enhanced heat transfer due to the presence of surface phonons \cite{Song2015}. These strongly modified emission and absorption properties may be favorably applied to the design of phononic metamaterials acting as extremely efficient passive coolers \cite{Rephaeli2013} or to the generation of improved detectors and emitters in the mid-IR, especially when surface-phonons interact with plasmons like in graphene-type materials \cite{Badioli2014}.

Beyond that, surface phonons concentrate electromagnetic energy at deep sub-wavelength scales in the same way as surface plasmons, but up to the  far-IR region, resulting in further intriguing nanophotonic applications  and techniques \cite{Yu2016}. SEIRA (surface enhanced infra-red absorption \cite{Le2008}) for instance, harnesses the strong and local enhancement of the infra-red electromagnetic field to probe molecular vibrations \cite{Neubrech2008}. This enhancement, usually ascribed to surface plasmons, can be significantly magnified by the exceptionally high-quality factors of surface phonons \cite{Neubrech2008,Autore2018}, leading to potential new approaches in biology \cite{NeunerIII2010}. Finally, metamaterials made up of surface phononic materials are key for super-resolution lenses in the mid-IR \cite{Caldwell2013}.
Strikingly, all these applications rely on the nanostructured electro-magnetic field in the vicinity of surfaces of metamaterials or nanoparticles. Designing or even engineering the electro-magnetic local density of states (EMLDOS) for specific functionalities require however the unambiguous visualization of such field modulations at the nanometer scale. This became accessible by near-field infra-red techniques \cite{Knoll1999,Hillenbrand2002,Zhang2013,DeWilde2006}, and more recently by electron energy loss spectroscopy (EELS) in a scanning transmission electron microscope (STEM) \cite{Krivanek2014,Lagos2017}. EELS can measure vibrations at the atom scale \cite{Zhang2013,venkatraman2019,Hage2020} and reveal dispersion relations of phonons \cite{Hage2018,Senga2019}. Nevertheless, intrinsic to those techniques, they only allow for two-dimensional imaging and do not provide directional field information from the start. Recently, in the visible range, tomographic tilting of plasmonic EELS data in combination with sophisticated eigenmode-based reconstruction algorithms have subsequently led to a series of publications \cite{Nicoletti2013,Collins2015,Hoerl2013a,Hoerl2015}, culminating in the full, three dimensional and vectorial reconstruction of plasmonic field at all frequencies in the visible regime \cite{Hoerl2017a,Haberfehlner2017}. The possibility to perform such a reconstruction is deeply rooted in the genuine relation between plasmon excitations and their EMLDOS \cite{Abajo2008a}, which describes the variation of the square modulus of the eigenfields, projected along arbitrary axes, in space and energy.
\begin{figure}[bthp]
    \centering
    \includegraphics[width= 0.8\columnwidth]{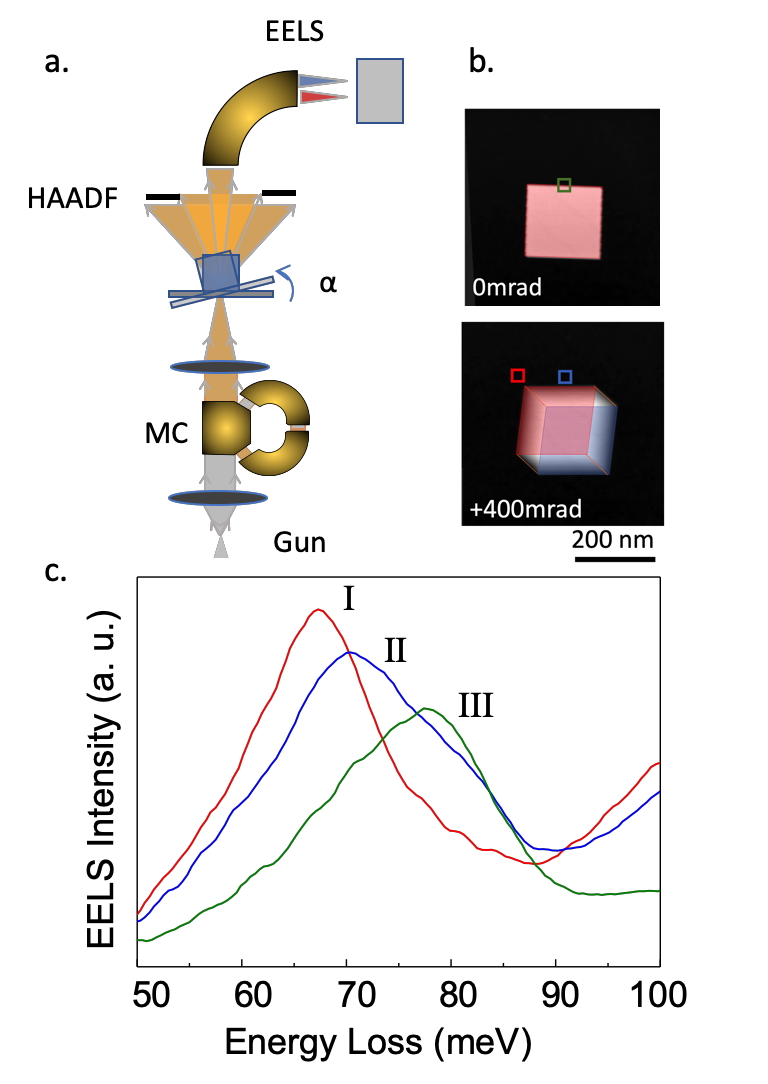}
    \caption{\textbf{Tomographic surface phonon EELS experiments}. a. Surface phonon tomography set-up. MC: Monochromator. b. High angle annular dark-field images of an MgO cube acquired at two different tilt angles. c. Selected spectra for the two different tilt angles taken at the positions indicated on b. Note the difference of spectra upon tilt for a fixed beam position (modes II and III). }
    \label{fig:Geometry}
\end{figure}

Given the strong formal analogy between surface plasmons and phonons \cite{Kliewer}, and having shown the theoretical link between EELS and surface phonon ELMDOS \cite{Lourenco-Martins2017}, it was therefore speculated whether a three-dimensional reconstruction of surface phonons properties is also feasible \cite{Lourenco-Martins2017}. The high demands for simultaneous nanometer spatial, meV spectral resolution and access to the infrared regime could recently be met, and further advancements in EELS tomographic reconstruction algorithms now paved the way for a full assessment of the 3D phononic ELMDOS. Here, we ally tilted EELS spectral-imaging with a model-constrained approach to give a comprehensive, 3D vectorial view on the EMLDOS of individual nanometer-scale MgO particles.

\begin{figure}[bthp]
    \includegraphics[width= \columnwidth]{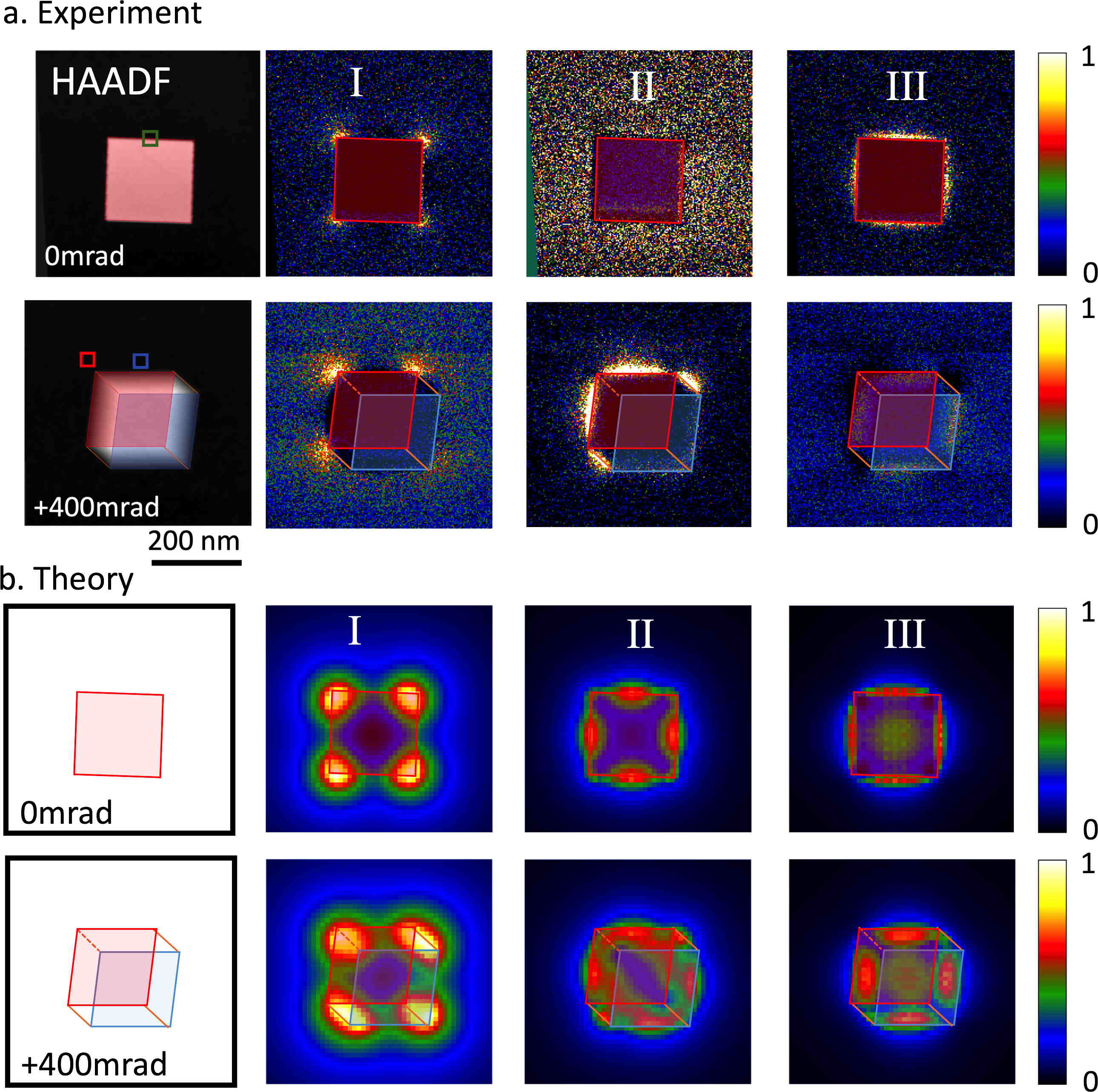}
    \caption{\textbf{2D phonon mapping at different tilt angles}. a. Experimental HAADF images (left) and filtered maps (right) extracted at energies of the main surface phonon modes (I, II, III) for two tilts configurations (0 and 400 mrd). The face in contact with the substrate is highlighted by a blue square. b. Same for simulations of a cube in vacuum.}
    \label{fig:Maps}
\end{figure}

We describe the experimental set-up in figure \ref{fig:Geometry}a. A 60 keV electron beam with an initial energy width of $\approx~350$ meV is filtered by a monochromator to obtain a final energy spread around $7$ to $10$ meV. This monochromator \cite{Krivanek2014}  efficiently optimizes  the current left after monochromation, with a beam current of a few pA in a sample area of $\approx~ 1$ nm$^2$. 
The nano-object presented in figure \ref{fig:Geometry} is a MgO cube with edge length of 191 nm deposited on a 20 nm thin $\mathrm{Si_3N_4}$  substrate. The back surface of the substrate was covered with a few nm thin carbon layer to avoid charge-related issues (see Materials and Methods). By scanning the electron beam, one can collect high angle annular dark field (HAADF) images that reveal the morphology of the cube. Sample tilting by an angle $\alpha$ allows imaging of the cube under different orientations (Figure \ref{fig:Geometry}b). At each position of the scan, an EELS spectrum was recorded. Figure \ref{fig:Geometry}c shows clear spectral responses in the upper far-IR, extracted at two different tilt angles  for two different positions of the electron beam. As the spectral features do change with both the electron beam position and the tilt angle, the selected  spectral, spatial and tilt resolutions of the used setup (see Materials and Methods) has proven adequate  to resolve directly signals characteristic of the main modes of cubes \cite{Lagos2017,Lourenco-Martins2017}. In order to understand their physical origin of the three main spectral features, numbered I, II and III,  we systematically recorded  EELS spectral-images (SI) at different tilt angles. For a first analysis, we present in figure \ref{fig:Maps}a  energy maps filtered at the energy maxima of the modes I, II and III, for two different tilt angles (see full tilt maps in the supplementary material). These maps are generated by using a fitting routine for each spectral mode of a SI (see Materials and Methods), and writing the resulting intensity in the filtered image pixel.  

The principal spatio-spectral features are already apparent. The MgO cube exhibits signals strongly localized in energy at 68 (mode I), 69 (mode II) and 78 meV (mode III). When the electron beam is propagating perpendicular to the faces that are parallel to the substrate (0 mrd tilt), mode I  is localized on the four corners. Tilted spectral-imaging directly shows a difference in intensity  between the signal  on corners in vacuum and those on the substrate. The latter is much weaker than the former. Mode II is not directly seen on the 0 mrd map, but becomes visible upon tilting (see also supplementary material). In this case, the signal  originates from the edges, and again the signal is much weaker for the edges attached to the substrate. Finally, the mode III is present at all angles. However, its spatial distribution is more difficult to understand. 

With the exception of mode I, which can be assigned to the corner, an unambiguous interpretation of the data is not  straightforward. In fact, the signal of mode II  is essentially revealed  for geometries where the beam direction is not parallel to an edge. Such an angular dependence points to the vectorial nature of the signal. Furthermore, the localization of the signal of mode III cannot be deduced directly from the maps. As presented in Figure \ref{fig:Maps}b and in the supplementary material, we performed simulations for a cube  of equal size in vacuum with the boundary element method (BEM) \cite{GarciadeAbajo2002} using the MNPBEM package \cite{Hohenester2014}. The simulations reveal similar features as the experimental data and confirm the existence of three different modes. The main deviations  are the absence of signal inside the cube in the experimental data and the absence of asymmetry in the theoretical data. The former relates to the  experimental issue of electrons getting scattered out of the spectrometer due to the large thickness. The latter is due to the fact that the substrate influences have not been accounted for in the simulations.

In order to understand the full surface phonons response of the cube, we used a tomographic reconstruction scheme.  Owing to the mathematical similarity between the surface plasmons and phonons, we adapted the methods recently developed to measure the plasmonic EMLDOS \cite{Hoerl2017} (see Materials and Methods and Theoretical approach in supplementary material) to the quasistatic limit. The quasistatic limit applies because the typical free space wavelengths of the surface phonons (several tens of $\mu m$) are much larger than the cube size. This leads to a simple description of the vibrational response in terms of well-defined eigenmodes with the following underlying principle. The vibrational response of the cube can be described as a set of modes indexed by a number $k$, ${\lambda_k,u_k(\vec{s})}$, where $\lambda_k$ is an eigenvalue, and $u_k(\vec{s})$ the spatial distribution of the associated eigencharge  ($\vec{s}$ being the surface coordinate) \cite{Lagos2017,Lourenco-Martins2017}.     From this set, all linear physical quantities can be deduced in this quasistatic limit \cite{GarciadeAbajo1997,Boudarham2012,Lourenco-Martins2017}. This includes  EELS and the EMLDOS, which are both linearly related to (see supplementary material):
\begin{equation}
    A(s,s')=\sum_k \{\frac{\lambda_k+\frac{1}{2}}{\Lambda(\omega)+\lambda_k}\}u_k(\vec{s})u_k(\vec{s'})
    \label{eq:assprime}
\end{equation}
where $\Lambda(\omega)$ is a material-dependent function that directly depends on the dielectric properties of the material at frequency $\omega$. The $\omega$ dependent prefactor determines a dispersion relation for the eigenvalues $\lambda_k$ so that a distinct spectral signature dominated by the surface phonon energy can be attributed to each mode \cite{Abajo1998,Lourenco-Martins2017}.

Due to this relation, a robust procedure for the reconstruction can  be envisaged. First, we measure EELS SIs for different tilt angles. Then  ${\lambda_k,u_k(\vec{s})}$ can be retrieved from the EELS data. This is done by fitting the EELS signal to its theoretical expression (see supplementary material) which is expressed as a function of equation \eqref{eq:assprime}.
To speed up the reconstruction, the sought eigenvectors $u_k(\vec{s})$ are projected on the eigenvector basis of an ideal structure $\tilde{u}_k(\vec{s})$. In the case of a cube, the ideal structure is a cube in vacuum, as presented in Figure \ref{fig:Maps}b. The eigenmodes of this struture are calculated using the boundary element method (BEM) \cite{GarciadeAbajo1997,Hohenester2012,Hohenester2014}, see supplementary material. Finally, at this point, $A(s,s')$ is known. Since the full EMLDOS is directly proportional to $A(s,s')$, the EMLDOS can then be deduced. The applicability of this procedure was validated on synthetic data based on a cube in vacuum.

\begin{figure*}[bthp]
    \includegraphics[width=\textwidth]{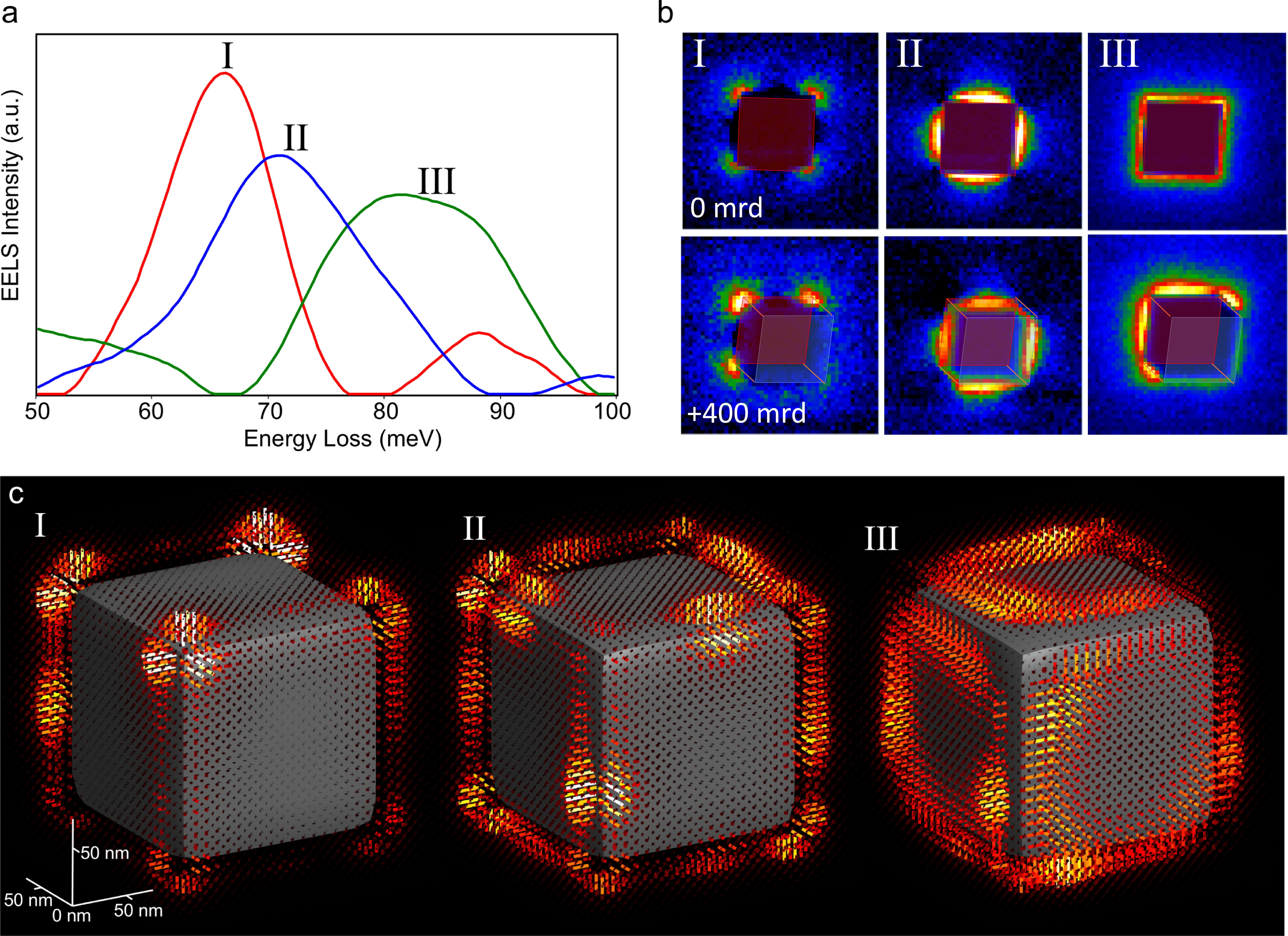}
    \caption{\textbf{Three dimensional, fully vectorial reconstruction of the phononic electromagnetic local density of state.} a. NMF component extracted from the experimental data. b. Reconstructed NMF maps for the 3 modes at the two angles shown in Fig. \ref{fig:Maps}. c. Three dimensional reconstruction of the EMLDOS seen from the top (the substrate, not shown, is at the bottom of the cube). The needles indicate the direction of the polarization of the EMLDOS, while the color indicates its intensity, from red to yellow. }
    \label{fig:Cube3D}
\end{figure*}

In order to perform the reconstruction, we used 12 spectral images (12 tilt angles), consisting in $400*400$ spectra (see Materials and Methods). Despite the massive amount of data, the signal to noise ratio was insufficient  to proceed directly to the reconstruction. Data treatment using the non-negative matrix factorization (NMF), following the mode extraction performed in the pioneer work on 3D non-vectorial surface plasmon reconstruction (\cite{Nicoletti2013} and Materials and Methods), leads to the  modal signatures as shown in Figure~\ref{fig:Cube3D}a. The NMF spectra are consisting of several peaks, with the most prominent ones corresponding to the I/II/III modes. The corresponding maps displayed in Figure~\ref{fig:Cube3D}b reproduce the spatial variation already observed on the raw data.
 These features are  fitted against the EELS expression as a function of $A(s,s')$  to reconstruct the surface phonon EMLDOS as displayed in Figure~\ref{fig:Cube3D}c, and shown for a variety of directions in the supplementary material (Figure~\ref{fig:orientations}). To assure data integrity, modelled re-projected 2D EELS maps calculated from the final reconstruction were directly compared to the experimental ones, and shows good agreement (see supplementary material).
With this reconstruction, mode I can be easily interpreted as a mode whose EMLDOS is mainly localized at the corner (”corner mode”), agreeing with previous experimental \cite{Lagos2017} and theoretical works \cite{Lourenco-Martins2017}. It gets severely damped close to the substrate, similarly to what happens for corner \textit{plasmonic} modes of cubes deposited on a surface \cite{Mazzucco2012}. The effect of the near-field environment modification on  the phonon properties, predicted in \cite{Lourenco-Martins2017} is revealed here for the first time. In passing, we note that, although the reconstruction basis is that of a perfectly symmetric  cube in vacuum, the final reconstruction, which combines several eigenvectors,, does nicely reproduce the asymmetry in the corner mode. In addition, some intensity can be seen near the center of some edges. This is because the reconstruction includes both resonant (corner-localized  \cite{Lourenco-Martins2017}) and non-resonant (edge-localized \cite{Lourenco-Martins2017})  contributions from the different modes. This points to the fact that the NMF does not provide a pure orthogonal decomposition of the different modes.  It is remarkable that our reconstruction scheme allows to observe this mixing although the EELS maps and the re-projection do not. This emphasizes the need of reconstructing a physical observable, the full EMLDOS, that contains the whole physical content of a peculiar system, in contrast to the EELS data, which cannot always be directly interpreted \cite{Hohenester2009}.    The mode II is  localized at the edges ("edge mode"), having been predicted of being a series of individual modes \cite{Lourenco-Martins2017}, which however could be neither  measured nor identified previously. The reason is twofold: the energy shift between the corner and edge mode is extremely small, and the \textit{vectorial} and spatial structure of the mode makes it difficult to probe it in the most common untilted projection geometry. Indeed, the electron only couples to eigenmodes if the electrical field is oriented along the beam direction. Also, experimental EELS signal is only accessible for non-penetrating geometries: the thickness of the cube results in a deflection of the electrons  away from the spectrometer entrance aperture, as commonly encountered for thick samples in EELS. We see  (see also supplementary figure 5) that the portion of the field coupled to the electron outside of the cube becomes increasingly higher as the tilt angle augments for mode II. The situation is opposite for mode III. This mode can be  related to surface modes (see fig. \ref{fig:Cube3D}c).  We finally note that the effect of the substrate leads to severe damping of the three modes in its vicinity, which we can attributed to the dissipation induced by the carbon layer. Supplementary data show that the damping is less severe in the absence of a carbon layer.

In conclusion, this first proof of principle visualization of the surface phonon EMLDOS should motivate the development of more systematic reconstruction of the full surface phonon optical response. Such mapping should be extended to other situations where the three dimensional and vectorial information of the electromagnetic density is of importance. This  includes extension of the methodology to anisotropic materials such as graphene analogues and transition metal dichalcogenides. This also includes the possibility to study consistently strong coupling physics, which has recently been unravelled for plasmons and phonons in EELS \cite{Tizei2020}. Finally, highly monochromated EELS has triggered much hope for its potential applications in vibrational mapping for biological systems \cite{Rez2016}. However, it is well-known that the 3D information is mandatory for this purpose \cite{Kukulski2012}. Therefore, the present method should be adapted to cryomicroscopy to e.g make it possible to combine   ultrastructure characterization with protein vibrational marking in 3D.

\

\section*{Acknowledgments}
XL and MK want to thanks M. Walls for help in preparing the sample.
This work has received support from the National Agency for Research
under the program of future investment TEMPOS-CHROMATEM with the Reference No. ANR-10-EQPX-50.
This project has received funding from the european union’s horizon 2020 research and innovation programme under grant agreement no 823717 and from the Austrian Science fund FWF under project P 31264

\clearpage
\newpage


\section*{Materials and Methods}
\subsection*{Sample preparation} 
\label{sub:sample_preparation}


Samples were prepared by collecting the fumes from an ignited strip of Mg on a standard TEM grid. We tried different TEM grids, lacey carbon grids (Agar, 300 mesh Cu grid), small $Si_3N_4$ membranes window (Ted Pella™, 9 windows 0.1*0.1mm with 15nm thick $Si_3N_4$), and big $Si_3N_4$ membrane window (TEMwindows, single 500 $\mu m$ window with 20nm thick Si3N4) back-covered with a thin (3-5 nm) layer of carbon. This last combination gives the best results, as it allows for a wide tilt angle without the shadow effect from the grid border, and eliminate the electrical charging related drift at high tilt angles (see supplementary). 
\subsection*{Experiments}
 Experiments were performed on a modified NION HERMES-200 S at 60 keV. The modification of the microscope includes a larger pole-piece gap (6 mm) which allows for almost arbitrary tilts. The EELS spectrometer is the NION-IRIS one, fitted with a Princeton instrument KURO sCMOS camera. We used 10 mrd incidence and 15 mrd collection angles. The current was a few pA and the spectral resolution in the 8-9 meV range. For each nanoparticle, a series of SI for typically 12 different angles are taken, with angles ranging from +1200 mrd to -1200 mrd ($\approx \pm 68^{\circ}$). Each spectral image was constituted by 400*400 pixels. The typical acquisition time for one spectrum was 8 ms, therefore the time for a SI was around 20 min. 

\subsection*{Data Analysis}

\textbf{Morphological reconstruction}: To reconstruct the morphology of the MgO cubes, the HAADF STEM images are employed. For reconstruction we assume the sample to be a perfect cube of unknown size, electron scattering power and orientation. To find  its size and orientation, model parameters are fitted to minimize the least-square error between all measured HAADF projections and projections of our model. The parameters we fit are: size of the cube, grey value (scattering power) of voxels within the cube, three angles determining the orientation of the cube and one angle determining the orientation of the tilt axis. For taking into account lateral shifts of projections (x- and y-direction) the maximum value of a filtered cross-correlation between measured and modeled projections is calculated within the optimization for each tilt angle in order to center projections. The final parameters provide the cube size, as well as its orientation with respect to all measured projections (knowing the tilt angles at which projections were acquired). These parameters are then used in the reconstruction of the phonon fields. Supplementary material shows a comparison between measured and modeled HAADF STEM projections. The fitted edge length of the cube is 191~nm.

\textbf{EELS Data Processing}: All EELS spectrum images were aligned by the maximum of the zero-loss peak in each spectrum and binned by 8x8, reducing the number of pixels of each spectrum image to a pixel size of 10~nm. After binning, all spectrum images were combined in a single dataset for further data processing. Before non-negative matrix factorization (NMF), all spectra were cropped to a range between 50~meV and 100~meV, which contains all major phonon peaks while excluding most of the zero-loss peak. The data was scaled to normalize the Poisson noise and an NMF algorithm \cite{Lin2007,Nicoletti2013} was applied for factorization using the Hyperspy software\cite{francisco_de_la_pena_2020_3973513}.

NMF was done for different numbers of components, where it was found that five components was a good choice. In this case three components could be attributed to surface phonon excitations, well localized in both energy and spatial location of the signals, while two components could be attributed to the tail of the zero-loss peak and other excitations (see Supplementary material for NMF factors and loadings at all tilt angles).
\subsection*{Simulation}
 In our simulation approach we compute the EEL spectra using the MNPBEM toolbox \cite{Hohenester2014}. For the dielectric function of MgO, we use a Lorentz oscillator model, as described in \cite{Lagos2017}. The experimentally determined cube size (191~nm) and orientation was used in the simulations. The substrate was not taken into account. Resonance energies were determined by simulating spectra at specific locations (corner, edge, face). This provided energies of 78~meV, 85~meV and 90~meV for the corner, edge and face mode. At these energies maps with the same pixel size and sample orientation as recorded in the experiment were simulated.
\subsection*{Reconstruction}
The main advantage of the present quasistatic approach compared to previous reconstruction schemes is that it authorizes to get the real eigenmodes as arbitrary hybridization of the ideal structure eigenmode, and strikingly to reconstruct the effect of the substrate, which is not explicitly considered in the evaluation of the basis (see figure 2b of the main text). 
On the other hand, this approach presents some limit for modes, like the face modes, which are genuinely highly degenerated (within the energy resolution of the experiment). Although for general geometry and modes, this situation is not common, the cube is a particularly complicated case, especially for the face modes \cite{Lourenco-Martins2017}, but even for the corner where the real eigenmode is essentially composed of two idealized eigenmodes. These degeneracies are usually lifted in the relativistic case, as they rely on quasi-normal modes computed for a given value of the energy $\hbar \omega$. In order to reproduce this degeneracy lift, a subset of modes is determined through energy filtering (see supplementary material, theoretical approach).

\section*{Supplementary text and figures}
\subsection*{Supplementary experimental data} 
\label{sub:supplementary_experimental_data}

\begin{figure*}[htb]
    \includegraphics[width= \textwidth]{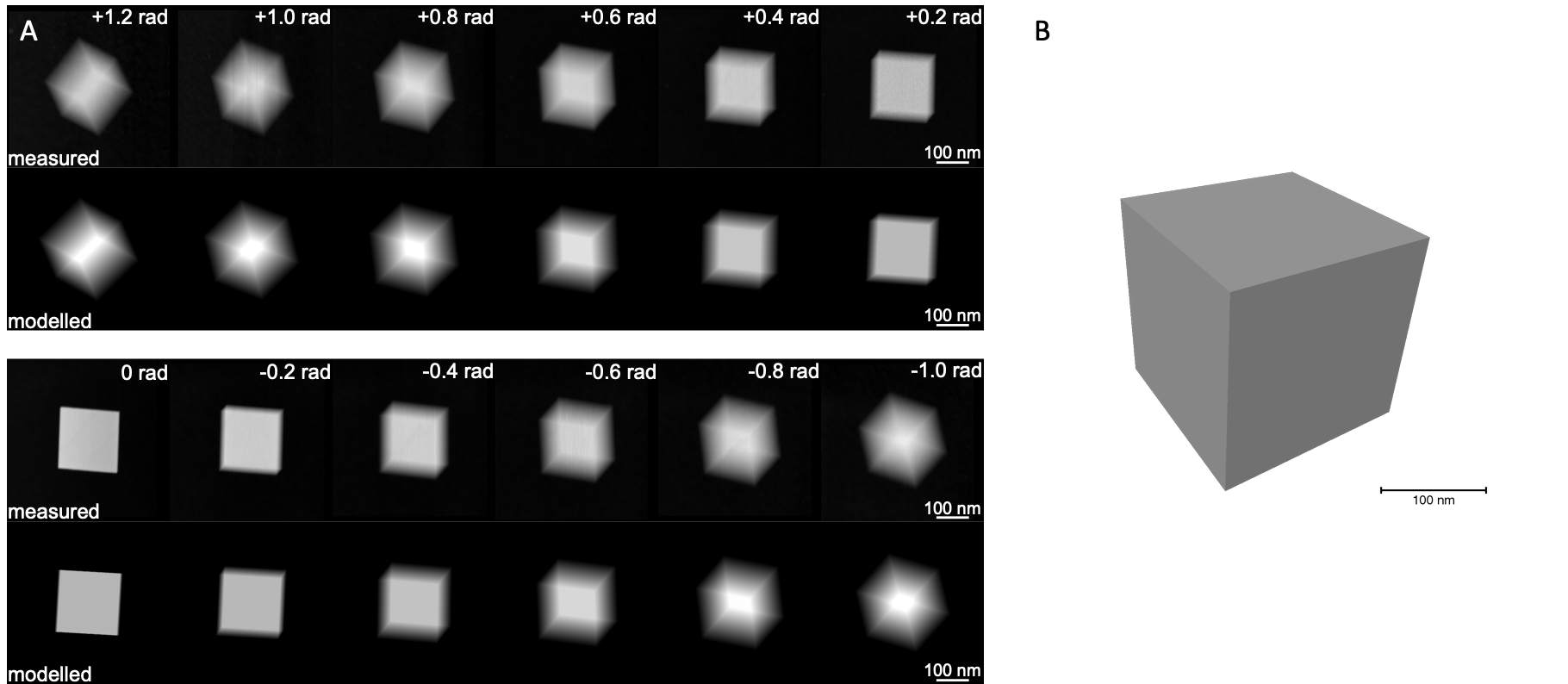}
    \caption{Reconstruction of the shape of the  MgO cube of the main text based on HAADF tilt series. A. comparison between the measured and modeled data. B. 3D reconstruction of the cube. }
    \label{fig:haadf}
\end{figure*}

\begin{figure*}[htb]
    \includegraphics[width= \textwidth]{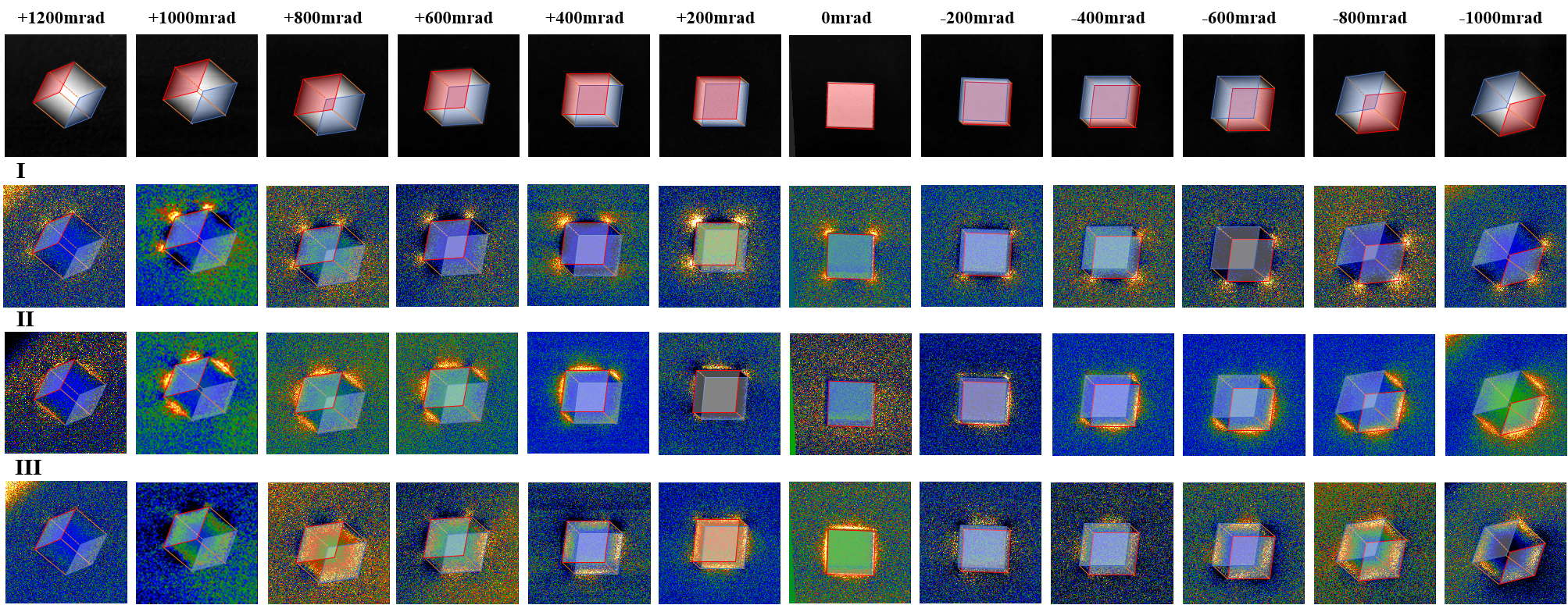}
    \caption{2D mapping at a series of tilt angles from +1200 mrad to -1000 mrad for the MgO cube of the main text. From top row to bottom, HAADF images at 12 different tilt angles, the corresponding filtered maps extracted at energies of the three main surface phonon modes. The color scale of each filtered map is adjusted independently to optimize the visualization of the signals.}
    \label{fig:exptilt}
\end{figure*}

\begin{figure*}[htb]
    \includegraphics[width= \textwidth]{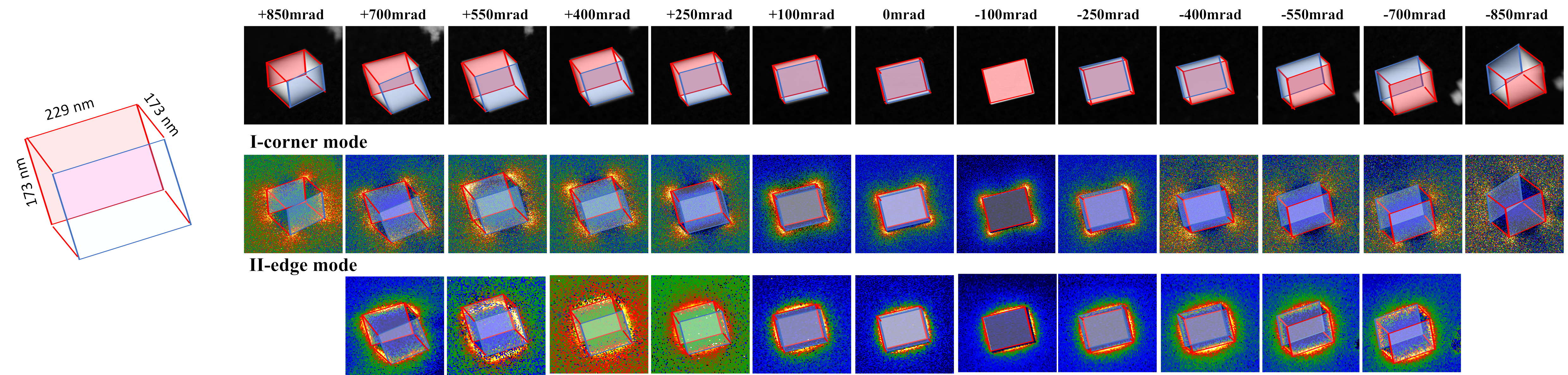}
    \caption{EELS spectral imaging of a tilt series of a parallelepiped MgO particle on a 15 nm thick Si3N4 membrane. The 2D phonon mappings of the MgO particle at a series of tilt angles from +850 mrad to -850 mrad are shown. From top row to bottom, HAADF images at 13 different tilt angles, the corresponding filtered maps extracted at energies of the two main surface phonon modes-corner and edge mode. The color scale of each filtered map is adjusted independently to optimize the visualization of the signals. No carbon has been deposited in that case, inducing a large drift at high tilt angles.}
    \label{fig:parallepiped}
\end{figure*}

We present the 3D reconstruction of the shape of the MgO cube presented in the main text in Figure \ref{fig:haadf}. Then, we show two different EELS spectral-imaging tilt series, the first one (Figure~\ref{fig:exptilt}) corresponds to the MgO cube presented in the main text, the second (Figure~\ref{fig:parallepiped}) is an example of an MgO parallelepiped deposited on a SiN substrate without carbon. In the later example, the absence of carbon induces charging that in turn induces too much drift at high tilt angles, so that the reconstruction becomes impossible.

\begin{figure*}[htb]
    \includegraphics[width= \textwidth]{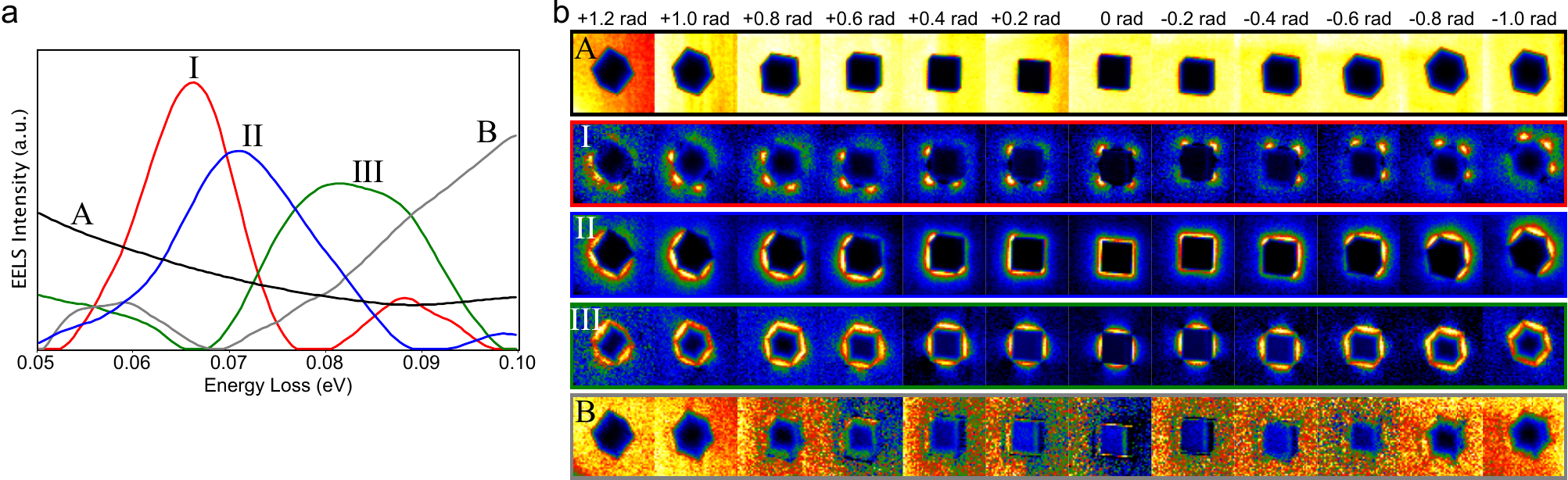}
    \caption{NMF factors (left) and loadings (right) as a function of tilt for the MgO cube of the main text used as a seed for the reconstruction.}
    \label{fig:loadings}
\end{figure*}

Finally,  in Figure \ref{fig:loadings} we present the NMF spectra and maps used as an input for the reconstruction of the EMLDOS of the cube in the main text.


\begin{figure*}[h]
    \centering
    \includegraphics[width= \textwidth]{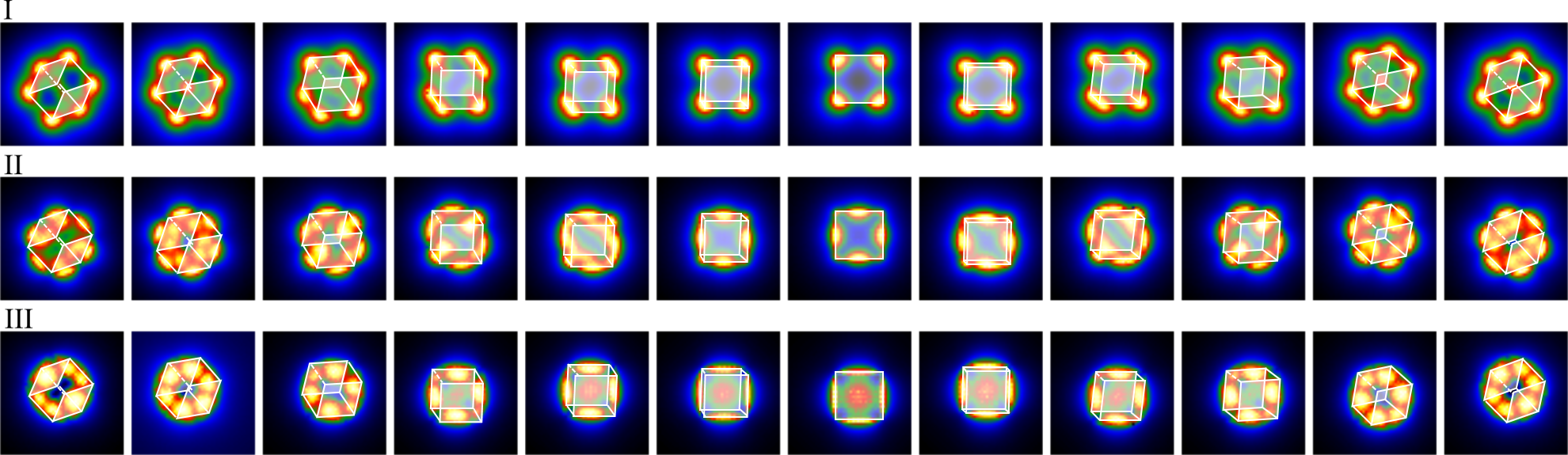}
    \caption{Simulated EELS maps for the three main modes for an MgO cube in vacuum.}
    \label{fig:SI_simulations}
\end{figure*}

\clearpage
\newpage
\subsection*{Theoretical approach}

Our theoretical approach for the \textsc{eels} tomography closely follows Refs.~\cite{hoerl:15,hoerl:17}, however, using the quasistatic approximation rather than the full Maxwell equations.  This approximation is well justified for the MgO cube with side lengths of approximately 191 nm and loss energies in the range between 20--100 meV.  In Refs.~\cite{hoerl:15,hoerl:17} we started from the resonance eigenmodes $\bm E_k(\bm r)$ for a given loss energy $\hbar\omega$, and approximated the dyadic Green's function (which completely characterizes the photonic environment) by a truncated set of these modes according to
\begin{equation}\label{eq:greendyad}
  \dy G(\bm r,\bm r')\approx\sum_{k=1}^n C_k\,\bm E_k(\bm r)\otimes\bm E_k(\bm r')\,.
\end{equation}
In a second step, the coefficients $C_k$ were obtained using a compressed-sensing optimization in order to get the best agreement between the experimental and reprojected \textsc{eels} maps, the latter being obtained using the dyadic Green's function of Eq.~\eqref{eq:greendyad}.

We next adapt the above scheme for the quasistatic approximation.  Our starting point is the boundary integral equation (we use SI units throughout)
\begin{equation}\label{eq:qsbim}
  {\rm \Lambda}(\omega)\sigma(\bm s)+\oint_{\partial\Omega}
  \frac{\partial G(\bm s,\bm s')}{\partial n}\sigma(\bm s')\,dS'=
  -\frac{\partial V_{\rm inc}(\bm s)}{\partial n}\,,
\end{equation}
which expresses the surface charge distribution $\sigma(\bm s)$, characterizing the response of the nanoparticle, in terms of the Green's function for the Poisson equation and the potential $V_{\rm inc}$, produced for instance by a swift electron or an oscillating dipole in the vicinity of the nanoparticle.  $\partial\Omega$ denotes the nanoparticle boundary, $\nicefrac\partial{\partial n}$ is the derivative in the direction of the outer surface normal, and $\Lambda(\omega)=\frac 12(\varepsilon_1+\varepsilon_2)/(\varepsilon_1-\varepsilon_2)$ is a quantity depending on the permittivities $\varepsilon_1$, $\varepsilon_2$ inside and outside the nanoparticle.  For further details see Refs.~\cite{boudarham:12,hohenester:20}.

A convenient way to solve the above equation is to seek for the eigenmodes $u_k(\bm s)$ and eigenvalues $\lambda_k$ of the integral operator on the left-hand side of Eq.~\eqref{eq:qsbim},
\begin{equation}\label{eq:eig1}
  \oint_{\partial\Omega}\frac{\partial G(\bm s,\bm s')}{\partial n}u_k(\bm s')\,dS' =
  \lambda_k\, u_k(\bm s)\,.
\end{equation}
It can be shown~\cite{ouyang:89} that $\lambda_k$ can be chosen real and the eigenmodes form a complete set such that any surface charge distribution can be expanded through these modes.  We can now express the probability that an electron loses in \textsc{eels} (aloof geometry) an energy $\hbar\omega$ via~\cite{boudarham:12,hohenester:20}
\begin{equation}\label{eq:eels1}
    \begin{split}
  \mathcal{P}_{\rm EELS}(\bm R_0,\omega)=&\frac{e^2}{\pi\hbar v^2\varepsilon_0}\sum_k\mbox{Im}
  \left\{\frac{\lambda_k+\frac 12}{{\rm\Lambda}(\omega)+\lambda_k}\right\}\\
  &\times \left|\oint_{\partial\Omega}V_{\rm el}(\bm s)u_k(\bm s)\,dS\right|^2\,,
  \end{split}
\end{equation}
where $V_{\rm el}(\bm r)$ is the electrostatic potential of the swift electron with impact parameter $\bm R_0$ and velocity $v$ (see below).  Similarly, the photonic local density of states (\textsc{EMLDOS}) projected along direction $\bm p$ is given by the expression~\cite{boudarham:12,hoerl:17,hohenester:20}
\begin{equation}\label{eq:ldos1}
    \begin{split}
  \rho_{\bm p}(\bm r)&=\rho_{\bm p}^{(0)}(\bm r)\\
  &+\frac 6\pi\sum_k\mbox{Im}\left\{\frac{\lambda_k+\frac 12}{{\rm\Lambda}(\omega)+\lambda_k}\right\}\left|\oint_{\partial\Omega}V_{\rm dip}(\bm s)u_k(\bm s)\,dS\right|^2\,,
  \end{split}
\end{equation}
where $\rho^{(0)}$ is the free-space density of states and $V_{\rm dip}$ is the potential of a dipole situated at position $\bm r$ with a unit dipole moment oriented along $\bm p$.  We next introduce
\begin{equation}\label{eq:tomo1}
  \mathcal{A}(\bm s,\bm s')=\sum_k\mbox{Im}
  \left\{\frac{\lambda_k+\frac 12}{{\rm\Lambda}(\omega)+\lambda_k}\right\}\,
  u_k(\bm s)u_k(\bm s')\,,
\end{equation}
which plays a similar role as the dyadic Green's function of Eq.~\eqref{eq:greendyad}, with the only exception that $\dy G$ additionally accounts for the propagation of the fields from the nanoparticle boundary to positions $\bm r$, $\bm r'$.  Apart from some constant prefactor, the loss probability of Eq.~\eqref{eq:eels1} can be written in the form
\begin{equation}\label{eq:eels2}
  \mathcal{P}_{\rm EELS}(\bm R_0,\omega)=const\times\oint V_{\rm el}(\bm s)\mathcal{A}(\bm s,\bm s')
  V_{\rm el}(\bm s')\,dSdS'\,.
\end{equation}
Similarly, the modification of the photonic \textsc{EMLDOS} of Eq.~\eqref{eq:ldos1} becomes
\begin{equation}\label{eq:ldos2}
  \rho_{\bm p}(\bm r)-\rho_{\bm p}^{(0)}(\bm r)=const\times\oint V_{\rm dip}(\bm s)\mathcal{A}(\bm s,\bm s')
  V_{\rm dip}(\bm s')\,dSdS'\,.
\end{equation}

In Refs.~\cite{hoerl:15,hoerl:17} we suggested a tomography scheme based on Eq.~\eqref{eq:greendyad}, where the eigenmodes $\bm E_k(\bm r)$ of a reference structure serve as a suitable basis for describing the experimental \textsc{eels} maps of a possibly modified particle geometry.  We can thus take the eigenmodes of the reference structure (for instance a perfect cube) and evaluate the \textsc{eels} maps for the experimental nanoparticle geometry (which might not be fully known) by optimizing the coefficients $C_k$.  A similar procedure can be applied in the quasistatic case.  We start by approximating Eq.~\eqref{eq:tomo1} with a truncated basis
\begin{equation}\label{eq:tomo2}
  \mathcal{A}(\bm s,\bm s')\approx\sum_{k=1}^nC_k\,
  u_k(\bm s)u_k(\bm s')\,,
\end{equation}
where the coefficients $C_k$ are again submitted to an optimization procedure in order to bring the reprojected \textsc{eels} maps of Eq.~\eqref{eq:eels2} as closely as possible to the experimental ones.  Eq.~\eqref{eq:tomo2} can accomodate for modifications of $\lambda_k$ and the permittivities $\varepsilon_1(\omega)$, $\varepsilon_2(\omega)$ between the reference and the experimental structure, however, it considers $u_k$ as the true eigenmodes for both the reference and the experimental structure.  For flat particle geometries with low symmetry this approximation works extremely well, as was shown in~\cite{hoerl:17}, but we here somewhat deviate from this approach (for reasons to be discussed further below) and also submit the eigenmodes to an additional optimization.  As $u_k$ provides a complete and orthogonal basis, we can obtain the true eigenmodes $\bar u_k$ through a transformation
\begin{equation}\label{eq:eigenmode1}
  \bar u_k(\bm s)\approx\sum_{k=1}^n \mathbb{Q}_{kk'}\,u_{k'}(\bm s)\,,
\end{equation}
where $\mathbb{Q}$ is an orthogonal matrix.  Eq.~\eqref{eq:eigenmode1} would be exact if we would consider all eigenmodes, but becomes an approximation for a truncated basis.  We next replace Eq.~\eqref{eq:tomo2} by the more general expression
\begin{equation}\label{eq:tomo3}
  \mathcal{A}(\bm s,\bm s')\approx\sum_{k=1}^n\sum_{k'=1}^n u_k(\bm s)\Big[\mathbb{Q}^T\mathbb{C}\mathbb{Q}\Big]_{kk'}
  u_{k'}(\bm s')\,,
\end{equation}
where $\mathbb{C}$ is a matrix with the coefficients $C_k$ on the diagonal.  Together with the matrix elements $\mathcal{V}_k=\oint V_{\rm el}(\bm s)u_k(\bm s)\,dS$ the loss probability of Eq.~\eqref{eq:eels2} can be cast to the form
\begin{equation}\label{eq:eels3}
  \mathcal{P}_{\rm EELS}(\bm R_0,\omega)\approx const\times\sum_{k=1}^n\sum_{k'=1}^n \mathcal{V}_k\Big[\mathbb{Q}^T\mathbb{C}\mathbb{Q}\Big]_{kk'}\mathcal{V}_{k'}\,.
\end{equation}
An expression similar to Eq.~\eqref{eq:eels3} can be also obtained for the \textsc{EMLDOS} modification of Eq.~\eqref{eq:ldos2}, with the only exception that $\mathcal{V}_k$ has to be computed for the dipole potential.

\begin{figure*}[t]
\centerline{\includegraphics[width=0.65\textwidth]{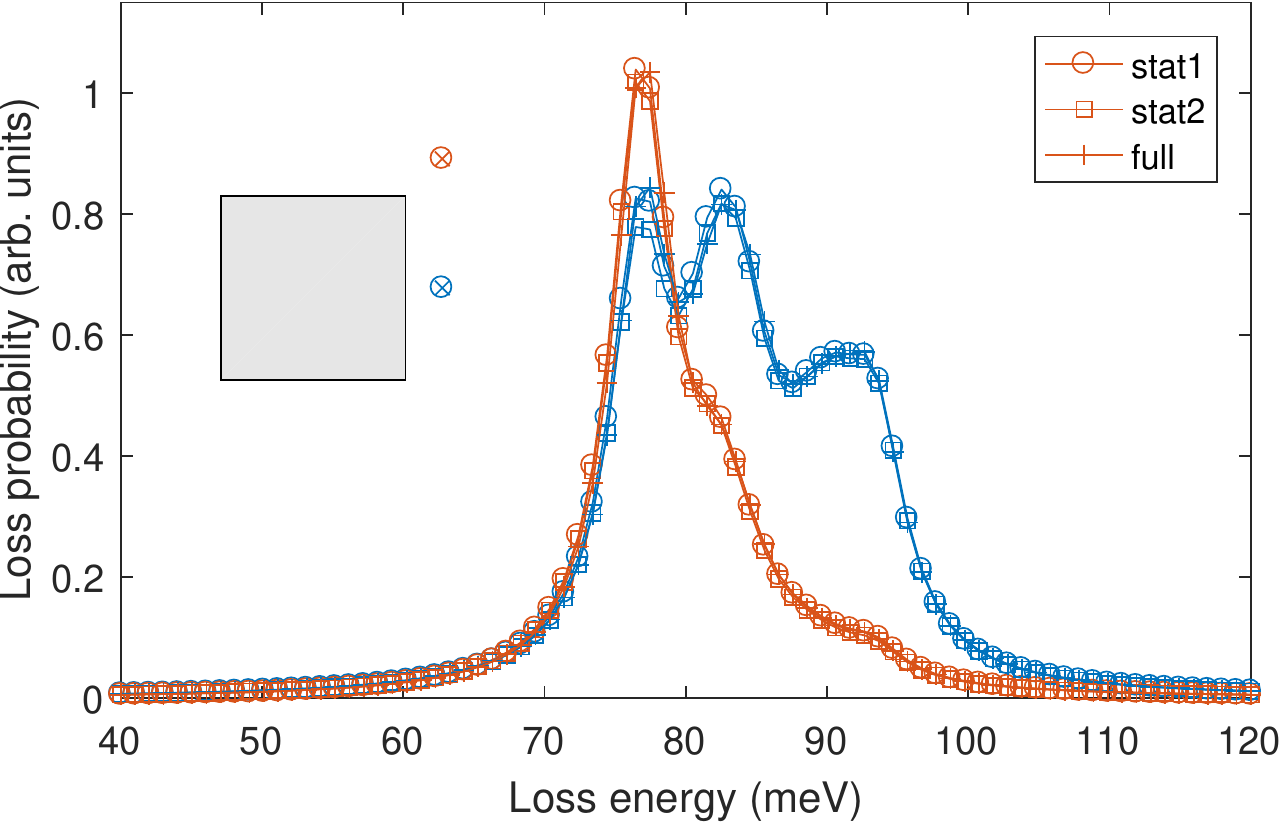}}
\caption{Quasistatic ($\circ$, $\square$) and full ($+$) simulation of \textsc{eels} probabilty for aloof electron trajectories with impact parameters shown in the inset.  Circles and squares show results for simulations using the full and small-distance approximations given in Eq.~\eqref{eq:elpotential}.  We use a MgO cube with a side length of 150 nm, a Drude-Lorentz permittivity for $\varepsilon_1$~\cite{hohenester.prb:18}, and $\varepsilon_2=1$.}
\label{fig:theory1}
\end{figure*}

The electrostatic potential of a swift electron propagating along $z$ is of the form~\cite{garcia:10}
\begin{equation}\label{eq:elpotential}
  V_{\rm el}(\bm r)=\left(\frac{e\omega}{2\pi v^2}\right)\,e^{iqz}K_0\left(q\rho\right)
  \underset{q\rho\ll 1}\longrightarrow
  -\left(\frac{e\omega}{2\pi v^2}\right)\,\ln(q\rho)\,,
\end{equation}
where $K_0$ is the modified Bessel function of order zero, $q=\nicefrac\omega v$, and $\rho$ is the distance between $(x,y)$ and the impact parameter $\bm R_0=(x_0,y_0)$.  The expression on the right-hand side is the expansion for small values of $q\rho$.  Fig.~\ref{fig:theory1} shows simulation results for the exact (circles)  and small-distance approximation (squares) for the electrostatic potential, showing almost identical results.  In our tomography approach we consider the more simple small-distance approximation where the $\ln\rho$ form corresponds to the potential of a charged wire, which is real-valued as we have already implicitly assumed in Eq.~\eqref{eq:eels3}.  In our tomography approach we express the orthogonal matrix in Eq.~\eqref{eq:tomo3} through $\mathbb{Q}=\exp(\mathbb{X})$, where $\mathbb{X}$ is a skew-symmetric matrix, and optimize the coefficients $C_k$ and the matrix elements of $\mathbb{X}$ using a least-squares minimization subject to the constraint that all $C_k$ elements must be positive or zero.

The eigenmodes $u_k$ are obtained from Eq.~\eqref{eq:eig1} using a Galerkin-type boundary element method (\textsc{bem}) approach~\cite{hohenester:20} where the cube boundary is approximated by triangles of finite size (in our simulations about 5000 triangles), and the surface charges are linearly interpolated from the vertices to the triangle interiors.  For the tomographic reconstruction of the corner and edge modes we take the forty eigenmodes with smallest eigenvalues $\lambda_k$.  Note that the mixing of eigenmodes, due to our use of Eq.~\eqref{eq:tomo3} rather than Eq.~\eqref{eq:tomo2}, was needed to obtain assymetric \textsc{EMLDOS} corner maps.  We found that all true eigenmodes $\bar u_k$ predominantly consist of the mixture between only two geometric eigenmodes $u_k$, and a proper symmetrization of $u_k$ would probably also suffice to obtain the same result.  However, we here stick to the more general approach based on an optimization of the orthogonal matrix~$\mathbb{Q}$.

For the face modes we are confronted with the problem that there exists a continuum of edge modes and the accumulation point of $\lambda_k$ for the edge modes lies below the $\lambda_k$ values for the face modes, such that in principle all eigenmodes must be considered in our tomographic approach, which renders it extremely difficult and computationally expensive.  Note that in our previous approach~\cite{hoerl:15,hoerl:17} based on Eq.~\eqref{eq:greendyad} this problem was absent because of the use of quasinormal modes, which are computed for a given loss energy $\hbar\omega$, and are therefore well adapted to the specific $\hbar\omega$ value---this is edge modes for the edge-mode loss energy and face-modes for the face-mode loss energy.  We here propose a similar scheme for the quasistatic modes, and compute at the face-mode loss energy the surface charge distribution $\sigma(\bm s,\bm r_\ell)$ for a collection of dipoles located a few tens of nanometers aways from the cube at positions $\bm r_\ell$.  The modes $u_k(\bm s)$ are then obtained from a singular value decomposition (\textsc{svd}) of
\begin{equation}\label{eq:svd}
    \begin{split}
  \sum_\ell\left(
  \mbox{Re}\Big[\sigma(\bm s,\bm r_\ell)\sigma(\bm s',\bm r_\ell)\Big]+
  \mbox{Im}\Big[\sigma(\bm s,\bm r_\ell)\sigma(\bm s',\bm r_\ell)\Big]\right)\\
  \approx\sum_{k=1}^n C_k u_k(\bm s)u_k(\bm s')\,,
  \end{split}
\end{equation}
thus providing us with the most important modes for the specific loss energy.  We found that the details of this approach are not overly sensitive to the number of dipoles and their respective positions, and we get converged results when including typically forty of these modes.  Note that the use of Eq.~\eqref{eq:svd} is somewhat specific for the cube geometry of our present concern, with the continuum of edge modes, whereas for other nanoparticle geometries we expect that a more unbiased choice depending on the $\lambda_k$ values should be sufficient.

\begin{figure*}
    \centering
\includegraphics[height=0.8\textheight]{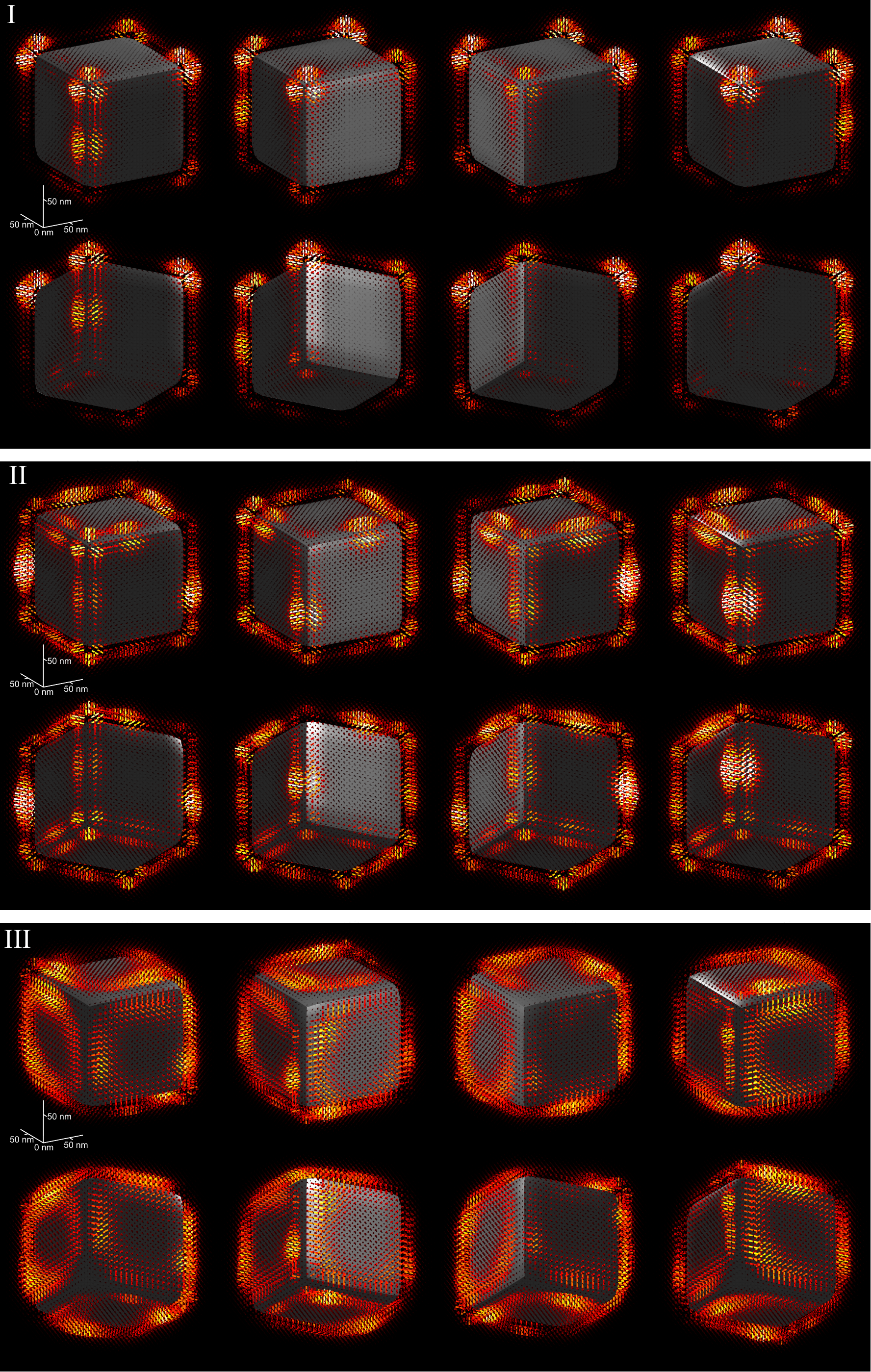}
\caption{Reconstructed vectorial \textsc{EMLDOS} for different orientations of the cube.}
\label{fig:orientations}
\end{figure*}

\begin{figure*}
\includegraphics[width=\textwidth]{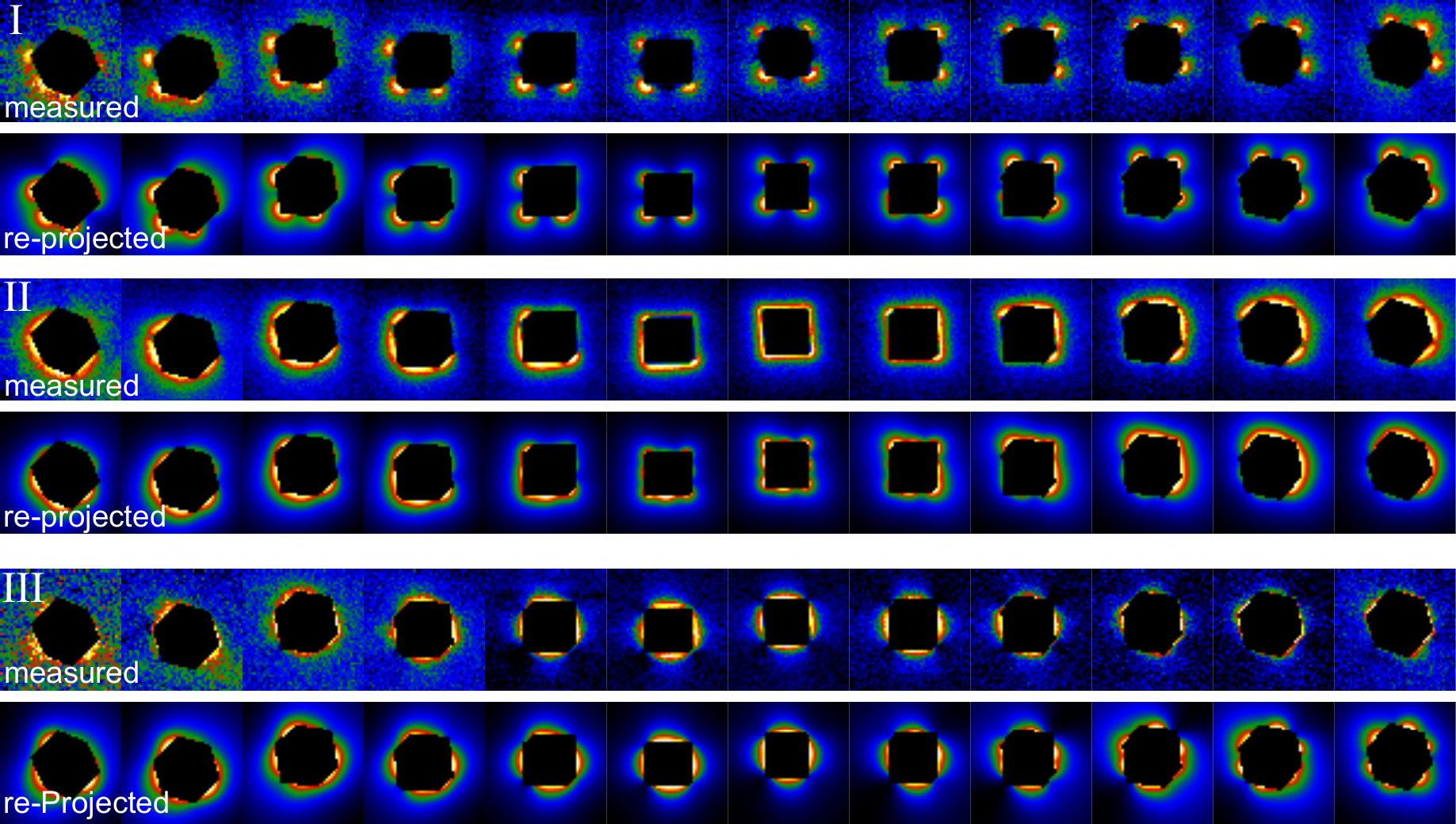}
\caption{Experimental and reprojected \textsc{eels} maps for corner (I), edge (Ii), and face (III) modes.}
\label{fig:retro}
\end{figure*}

\begin{figure*}
    \centering
\includegraphics[height=0.8\textheight]{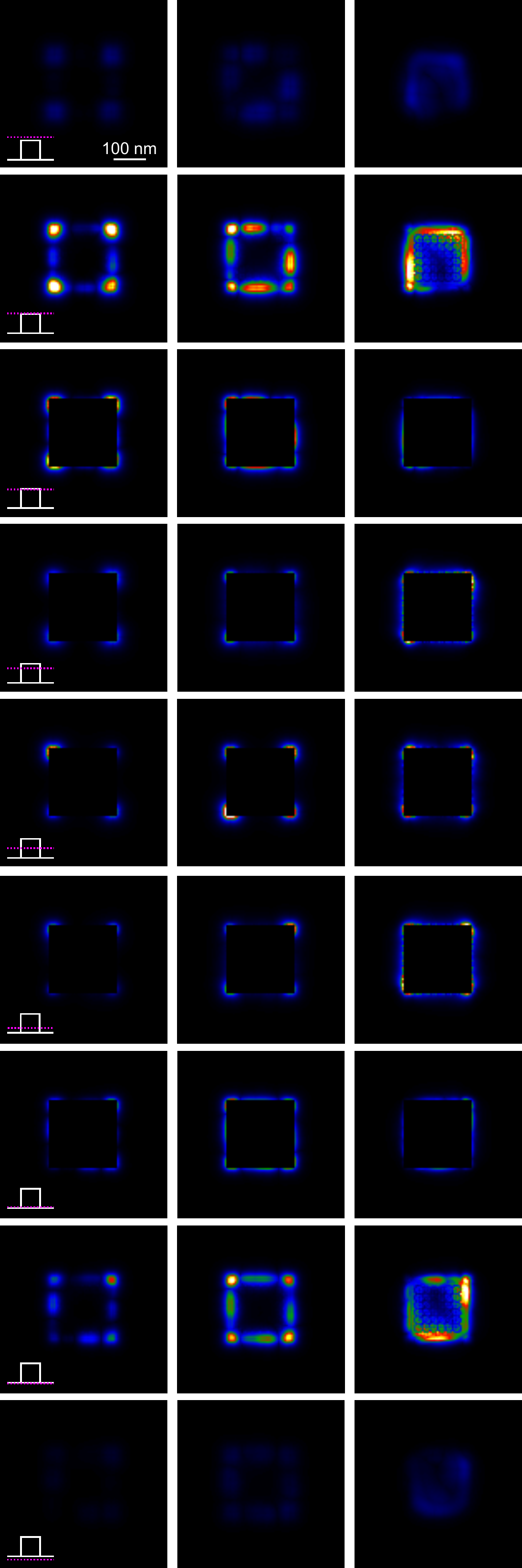}
\caption{Reconstructed total \textsc{EMLDOS} maps in different planes for corner (left), edge (middle), and face (right) modes. In the schematics, the position of the slicing plane is indicated with a dotted line, and the substrate by a line.}
\label{fig:slice}
\end{figure*}
The reconstruction for several angle of views is presented in Figure~\ref{fig:orientations}. From this reconstruction, re-projection can be generated and compared to the original data (Figure~\ref{fig:retro}). Finally, the EMLDOS can be analyzed more precisely by slicing it (see Figure~\ref{fig:slice}).

\clearpage
\newpage

\bibliography{ThreeDPhonons,biblio}

\end{document}